\documentclass[11pt,a4paper]{article}

\usepackage{fullpage}
\usepackage{multirow}
\usepackage{graphicx}
\usepackage{multicol}
\usepackage{fancyvrb}
\usepackage{color}

\usepackage{listings}
\newcommand{\programname}[1]{\textbf{#1}\hspace{0.1in}}

\makeatletter
\def\PY@reset{\let\PY@it=\relax \let\PY@bf=\relax%
    \let\PY@ul=\relax \let\PY@tc=\relax%
    \let\PY@bc=\relax \let\PY@ff=\relax}
\def\PY@tok#1{\csname PY@tok@#1\endcsname}
\def\PY@toks#1+{\ifx\relax#1\empty\else%
    \PY@tok{#1}\expandafter\PY@toks\fi}
\def\PY@do#1{\PY@bc{\PY@tc{\PY@ul{%
    \PY@it{\PY@bf{\PY@ff{#1}}}}}}}
\def\PY#1#2{\PY@reset\PY@toks#1+\relax+\PY@do{#2}}

\expandafter\def\csname PY@tok@gd\endcsname{\def\PY@tc##1{\textcolor[rgb]{0.63,0.00,0.00}{##1}}}
\expandafter\def\csname PY@tok@gu\endcsname{\let\PY@bf=\textbf\def\PY@tc##1{\textcolor[rgb]{0.50,0.00,0.50}{##1}}}
\expandafter\def\csname PY@tok@gt\endcsname{\def\PY@tc##1{\textcolor[rgb]{0.00,0.27,0.87}{##1}}}
\expandafter\def\csname PY@tok@gs\endcsname{\let\PY@bf=\textbf}
\expandafter\def\csname PY@tok@gr\endcsname{\def\PY@tc##1{\textcolor[rgb]{1.00,0.00,0.00}{##1}}}
\expandafter\def\csname PY@tok@cm\endcsname{\def\PY@tc##1{\textcolor[rgb]{0.53,0.53,0.53}{##1}}}
\expandafter\def\csname PY@tok@vg\endcsname{\let\PY@bf=\textbf\def\PY@tc##1{\textcolor[rgb]{0.87,0.47,0.00}{##1}}}
\expandafter\def\csname PY@tok@m\endcsname{\let\PY@bf=\textbf\def\PY@tc##1{\textcolor[rgb]{0.40,0.00,0.93}{##1}}}
\expandafter\def\csname PY@tok@mh\endcsname{\let\PY@bf=\textbf\def\PY@tc##1{\textcolor[rgb]{0.00,0.33,0.53}{##1}}}
\expandafter\def\csname PY@tok@cs\endcsname{\let\PY@bf=\textbf\def\PY@tc##1{\textcolor[rgb]{0.80,0.00,0.00}{##1}}}
\expandafter\def\csname PY@tok@ge\endcsname{\let\PY@it=\textit}
\expandafter\def\csname PY@tok@vc\endcsname{\def\PY@tc##1{\textcolor[rgb]{0.20,0.40,0.60}{##1}}}
\expandafter\def\csname PY@tok@il\endcsname{\let\PY@bf=\textbf\def\PY@tc##1{\textcolor[rgb]{0.00,0.00,0.87}{##1}}}
\expandafter\def\csname PY@tok@go\endcsname{\def\PY@tc##1{\textcolor[rgb]{0.53,0.53,0.53}{##1}}}
\expandafter\def\csname PY@tok@cp\endcsname{\def\PY@tc##1{\textcolor[rgb]{0.33,0.47,0.60}{##1}}}
\expandafter\def\csname PY@tok@gi\endcsname{\def\PY@tc##1{\textcolor[rgb]{0.00,0.63,0.00}{##1}}}
\expandafter\def\csname PY@tok@gh\endcsname{\let\PY@bf=\textbf\def\PY@tc##1{\textcolor[rgb]{0.00,0.00,0.50}{##1}}}
\expandafter\def\csname PY@tok@ni\endcsname{\let\PY@bf=\textbf\def\PY@tc##1{\textcolor[rgb]{0.53,0.00,0.00}{##1}}}
\expandafter\def\csname PY@tok@nl\endcsname{\let\PY@bf=\textbf\def\PY@tc##1{\textcolor[rgb]{0.60,0.47,0.00}{##1}}}
\expandafter\def\csname PY@tok@nn\endcsname{\let\PY@bf=\textbf\def\PY@tc##1{\textcolor[rgb]{0.05,0.52,0.71}{##1}}}
\expandafter\def\csname PY@tok@no\endcsname{\let\PY@bf=\textbf\def\PY@tc##1{\textcolor[rgb]{0.00,0.20,0.40}{##1}}}
\expandafter\def\csname PY@tok@na\endcsname{\def\PY@tc##1{\textcolor[rgb]{0.00,0.00,0.80}{##1}}}
\expandafter\def\csname PY@tok@nb\endcsname{\def\PY@tc##1{\textcolor[rgb]{0.00,0.44,0.13}{##1}}}
\expandafter\def\csname PY@tok@nc\endcsname{\let\PY@bf=\textbf\def\PY@tc##1{\textcolor[rgb]{0.73,0.00,0.40}{##1}}}
\expandafter\def\csname PY@tok@nd\endcsname{\let\PY@bf=\textbf\def\PY@tc##1{\textcolor[rgb]{0.33,0.33,0.33}{##1}}}
\expandafter\def\csname PY@tok@ne\endcsname{\let\PY@bf=\textbf\def\PY@tc##1{\textcolor[rgb]{1.00,0.00,0.00}{##1}}}
\expandafter\def\csname PY@tok@nf\endcsname{\let\PY@bf=\textbf\def\PY@tc##1{\textcolor[rgb]{0.00,0.40,0.73}{##1}}}
\expandafter\def\csname PY@tok@si\endcsname{\def\PY@bc##1{\setlength{\fboxsep}{0pt}\colorbox[rgb]{0.93,0.93,0.93}{\strut ##1}}}
\expandafter\def\csname PY@tok@s2\endcsname{\def\PY@bc##1{\setlength{\fboxsep}{0pt}\colorbox[rgb]{1.00,0.94,0.94}{\strut ##1}}}
\expandafter\def\csname PY@tok@vi\endcsname{\def\PY@tc##1{\textcolor[rgb]{0.20,0.20,0.73}{##1}}}
\expandafter\def\csname PY@tok@nt\endcsname{\def\PY@tc##1{\textcolor[rgb]{0.00,0.47,0.00}{##1}}}
\expandafter\def\csname PY@tok@nv\endcsname{\def\PY@tc##1{\textcolor[rgb]{0.60,0.40,0.20}{##1}}}
\expandafter\def\csname PY@tok@s1\endcsname{\def\PY@bc##1{\setlength{\fboxsep}{0pt}\colorbox[rgb]{1.00,0.94,0.94}{\strut ##1}}}
\expandafter\def\csname PY@tok@gp\endcsname{\let\PY@bf=\textbf\def\PY@tc##1{\textcolor[rgb]{0.78,0.36,0.04}{##1}}}
\expandafter\def\csname PY@tok@sh\endcsname{\def\PY@bc##1{\setlength{\fboxsep}{0pt}\colorbox[rgb]{1.00,0.94,0.94}{\strut ##1}}}
\expandafter\def\csname PY@tok@ow\endcsname{\let\PY@bf=\textbf\def\PY@tc##1{\textcolor[rgb]{0.00,0.00,0.00}{##1}}}
\expandafter\def\csname PY@tok@sx\endcsname{\def\PY@tc##1{\textcolor[rgb]{0.87,0.13,0.00}{##1}}\def\PY@bc##1{\setlength{\fboxsep}{0pt}\colorbox[rgb]{1.00,0.94,0.94}{\strut ##1}}}
\expandafter\def\csname PY@tok@bp\endcsname{\def\PY@tc##1{\textcolor[rgb]{0.00,0.44,0.13}{##1}}}
\expandafter\def\csname PY@tok@c1\endcsname{\def\PY@tc##1{\textcolor[rgb]{0.53,0.53,0.53}{##1}}}
\expandafter\def\csname PY@tok@kc\endcsname{\let\PY@bf=\textbf\def\PY@tc##1{\textcolor[rgb]{0.00,0.53,0.00}{##1}}}
\expandafter\def\csname PY@tok@c\endcsname{\def\PY@tc##1{\textcolor[rgb]{0.53,0.53,0.53}{##1}}}
\expandafter\def\csname PY@tok@mf\endcsname{\let\PY@bf=\textbf\def\PY@tc##1{\textcolor[rgb]{0.40,0.00,0.93}{##1}}}
\expandafter\def\csname PY@tok@err\endcsname{\def\PY@tc##1{\textcolor[rgb]{1.00,0.00,0.00}{##1}}\def\PY@bc##1{\setlength{\fboxsep}{0pt}\colorbox[rgb]{1.00,0.67,0.67}{\strut ##1}}}
\expandafter\def\csname PY@tok@kd\endcsname{\let\PY@bf=\textbf\def\PY@tc##1{\textcolor[rgb]{0.00,0.53,0.00}{##1}}}
\expandafter\def\csname PY@tok@ss\endcsname{\def\PY@tc##1{\textcolor[rgb]{0.67,0.40,0.00}{##1}}}
\expandafter\def\csname PY@tok@sr\endcsname{\def\PY@tc##1{\textcolor[rgb]{0.00,0.00,0.00}{##1}}\def\PY@bc##1{\setlength{\fboxsep}{0pt}\colorbox[rgb]{1.00,0.94,1.00}{\strut ##1}}}
\expandafter\def\csname PY@tok@mo\endcsname{\let\PY@bf=\textbf\def\PY@tc##1{\textcolor[rgb]{0.27,0.00,0.93}{##1}}}
\expandafter\def\csname PY@tok@mi\endcsname{\let\PY@bf=\textbf\def\PY@tc##1{\textcolor[rgb]{0.00,0.00,0.87}{##1}}}
\expandafter\def\csname PY@tok@kn\endcsname{\let\PY@bf=\textbf\def\PY@tc##1{\textcolor[rgb]{0.00,0.53,0.00}{##1}}}
\expandafter\def\csname PY@tok@o\endcsname{\def\PY@tc##1{\textcolor[rgb]{0.20,0.20,0.20}{##1}}}
\expandafter\def\csname PY@tok@kr\endcsname{\let\PY@bf=\textbf\def\PY@tc##1{\textcolor[rgb]{0.00,0.53,0.00}{##1}}}
\expandafter\def\csname PY@tok@s\endcsname{\def\PY@bc##1{\setlength{\fboxsep}{0pt}\colorbox[rgb]{1.00,0.94,0.94}{\strut ##1}}}
\expandafter\def\csname PY@tok@kp\endcsname{\let\PY@bf=\textbf\def\PY@tc##1{\textcolor[rgb]{0.00,0.20,0.53}{##1}}}
\expandafter\def\csname PY@tok@w\endcsname{\def\PY@tc##1{\textcolor[rgb]{0.73,0.73,0.73}{##1}}}
\expandafter\def\csname PY@tok@kt\endcsname{\let\PY@bf=\textbf\def\PY@tc##1{\textcolor[rgb]{0.20,0.20,0.60}{##1}}}
\expandafter\def\csname PY@tok@sc\endcsname{\def\PY@tc##1{\textcolor[rgb]{0.00,0.27,0.87}{##1}}}
\expandafter\def\csname PY@tok@sb\endcsname{\def\PY@bc##1{\setlength{\fboxsep}{0pt}\colorbox[rgb]{1.00,0.94,0.94}{\strut ##1}}}
\expandafter\def\csname PY@tok@k\endcsname{\let\PY@bf=\textbf\def\PY@tc##1{\textcolor[rgb]{0.00,0.53,0.00}{##1}}}
\expandafter\def\csname PY@tok@se\endcsname{\let\PY@bf=\textbf\def\PY@tc##1{\textcolor[rgb]{0.40,0.40,0.40}{##1}}\def\PY@bc##1{\setlength{\fboxsep}{0pt}\colorbox[rgb]{1.00,0.94,0.94}{\strut ##1}}}
\expandafter\def\csname PY@tok@sd\endcsname{\def\PY@tc##1{\textcolor[rgb]{0.87,0.27,0.13}{##1}}}

\def\PYZus{\char`\_}

\def\PYZlt{\char`\<}
\def\PYZgt{\char`\>}

\def\PYZhy{\char`\-}

\def\PYZdq{\char`\"}

\makeatother

\newcommand{\numParticipants}{162}
\newcommand{\numMale}{129}
\newcommand{\numFemale}{30}
\newcommand{\numGenderless}{3}
\newcommand{\numBloomington}{29}
\newcommand{\numMturk}{130}
\newcommand{\numOnline}{3}

\newcommand{\numTrials}{1,602}
\newcommand{\numPerfectTrials}{1,007}

\newcommand{\numPyBeginners}{94}

\newcommand{\minRT}{14}
\newcommand{\maxRT}{256}
\newcommand{\numOutliers}{60}

\newcommand{\expStartDate}{November 20, 2012}
\newcommand{\expEndDate}{January 19, 2013}

\newcommand{\meanAge}{28.4}
\newcommand{\meanProgYears}{6.9}
\newcommand{\meanPyYears}{2.0}
\newcommand{\percentDegree}{69.8}
\newcommand{\percentCS}{52.5}
\title{What Makes Code Hard to Understand?}
\author{{\bf Michael Hansen (\texttt{mihansen@indiana.edu})} \\
  School of Informatics and Computing, 2719 E. 10th Street \\
  Bloomington, IN 47408 USA
  \and {\bf Robert L. Goldstone (\texttt{rgoldsto@indiana.edu})} \\
  Dept.\ of Psychological and Brain Sciences, 1101 E. 10th Street \\
  Bloomington, IN 47405 USA
  \and {\bf Andrew Lumsdaine (\texttt{lums@indiana.edu})} \\
  School of Informatics and Computing, 2719 E. 10th Street \\
  Bloomington, IN 47408 USA}

\begin{document}
\maketitle

\begin{abstract}

What factors impact the comprehensibility of code? Previous research suggests
that expectation-congruent programs should take less time to understand and be
less prone to errors.  We present an experiment in which participants with
programming experience predict the exact output of ten small Python programs. We
use subtle differences between program versions to demonstrate that seemingly
insignificant notational changes can have profound effects on correctness and
response times. Our results show that experience increases performance in most
cases, but may hurt performance significantly when underlying assumptions about
related code statements are violated.

\end{abstract}

\section{Introduction}

The design, creation and interpretation of computer programs are some of the
most cognitively challenging tasks that humans perform.  Understanding the
factors that impact the cognitive complexity of code is important for both
applied and theoretical reasoning.  Practically, an enormous amount of time is
spent developing programs, and even more time is spent debugging them, and so if
we can identify factors that expedite these activities, a large amount of time
and money can be saved.  Theoretically, programming is an excellent task for
studying representation, working memory, planning, and problem solving in the
real world.

We present a web-based experiment in which participants with a wide variety of
Python and overall programming experience predict the output of ten small Python
programs.  Most of the program texts are less than 20 lines long and have fewer
than 8 linearly independent paths (known as cyclomatic
complexity~\cite{Mccabe1976}). Each program \emph{type} has two or three
\emph{versions} with subtle differences that do not significantly change their
lines of code (LOC) or cyclomatic complexities (CC).  For each participant and
program, we grade text responses on a 10-point scale, and record the amount of
time taken.  The different versions of our programs were designed to test a
couple of underlying questions.  First, ``How are programmers affected by
programs that violate their expectations, and does this vary with expertise?''
Previous research suggests that programs that violate expectations should take
longer to process and be more error-prone than expectation-congruent programs.
There are reasons to expect this benefit for expectation-congruency to interact
with experience in opposing ways. Experienced programmers may show a larger
influence of expectations due to prolonged training, but they may also have more
untapped cognitive resources available for monitoring expectation violations. In
fact, given the large percentage of programming time that involves debugging (it
is a common saying that 90\% of development time is spent debugging 10\% of the
code), experienced programmers may have developed dedicated monitors for certain
kinds of expectation-violating code.

The second question is: ``How are programmers influenced by physical
characteristics of notation, and does this vary with expertise?''  Programmers
often feel like the physical properties of notation have only a minor influence
on their interpretation process.  When in a hurry, they frequently dispense with
recommended variable naming, indentation, and formatting as superficial and
inconsequential.  However, in other formal reasoning domains such as
math~\cite{goldstone2010education}, apparently superficial formatting influences
such as physical spacing between operators has been shown to have a profound
impact on performance.  Furthermore, there is an open question as to whether
experienced or inexperienced programmers are more influenced by these physical
aspects of code notation.  Experienced programmers may show less influence of
these ``superficial'' aspects because they are responding to the deep structure
of the code.  By contrast, in math reasoning, experienced individuals sometimes
show more influence of notational properties of the symbols, apparently because
they use perception-action shortcuts involving these properties in order to
attain efficiency~\cite{goldstone2010education}.

\section{Related Work}

Psychologists have been studying programmers for at least forty years.  Early
research focused on correlations between task performance and human/language
factors, such as how the presence of code comments impacts scores on a program
comprehension questionnaire. More recent research has revolved around the
cognitive processes underlying program comprehension.  Effects of expertise,
task, and available tools on program understanding have been
found~\cite{Detienne2002}. Studies with experienced programmers have revealed
conventions, or ``rules of discourse,'' that can have a profound impact
(sometimes negative) on expert program
comprehension~\cite{soloway1984empirical}.

Our present research focuses on programs much less complicated than those the
average professional programmer typically encounters on a daily basis. The
demands of our task are still high, however, because participants must predict
precise program output. In this way, it is similar to debugging a short snippet
of a larger program. Code studies often take the form of a code review, where
programmers must locate errors or answer comprehension questions after the fact
(e.g., does the program define a Professor class?~\cite{burkhardt2002object}).
Our task differs by asking programmers to mentally simulate code without
necessarily understanding its purpose. In most programs, we intentionally use
meaningless identifier names where appropriate (variables \texttt{a},
\texttt{b}, etc.) to avoid influencing the programmer's mental model. 

Similar research has asked beginning (CS1) programming students to read and
write code with simple goals, such as the Rainfall
Problem~\cite{Guzdial2011_science}. To solve it, students must write a program
that averages a list of numbers (rainfall amounts), where the list is terminated
with a specific value -- e.g., a negative number or 999999. CS1 students perform
poorly on the Rainfall Problem across institutions around the world, inspiring
researchers to seek better teaching methods. Our work includes many Python
novices with a year or less of experience (\numPyBeginners~out of
\numParticipants), so our results may contribute to ongoing research in early
programming education. 

\section{Methods}

One hundred and sixy-two participants (\numMale~males, \numFemale~females,
\numGenderless~unreported) were recruited from the Bloomington, IN area
(\numBloomington), on Amazon's Mechanical Turk (\numMturk), and via e-mail
(\numOnline). All participants were required to have some experience with
Python, though we welcomed beginners. The mean age was \meanAge~years, with an
average of \meanPyYears~years of self-reported Python experience and
\meanProgYears~years of programming experience overall.  Most of the
participants had a college degree (\percentDegree\%), and were current or former
Computer Science majors (\percentCS\%). Participants from Bloomington were paid
\$10, and performed the experiment in front of an eye-tracker (see Future Work).
Mechanical Turk participants were paid \$0.75.

\begin{figure}[ht]
  \begin{center}
    \includegraphics[width=0.85\textwidth]{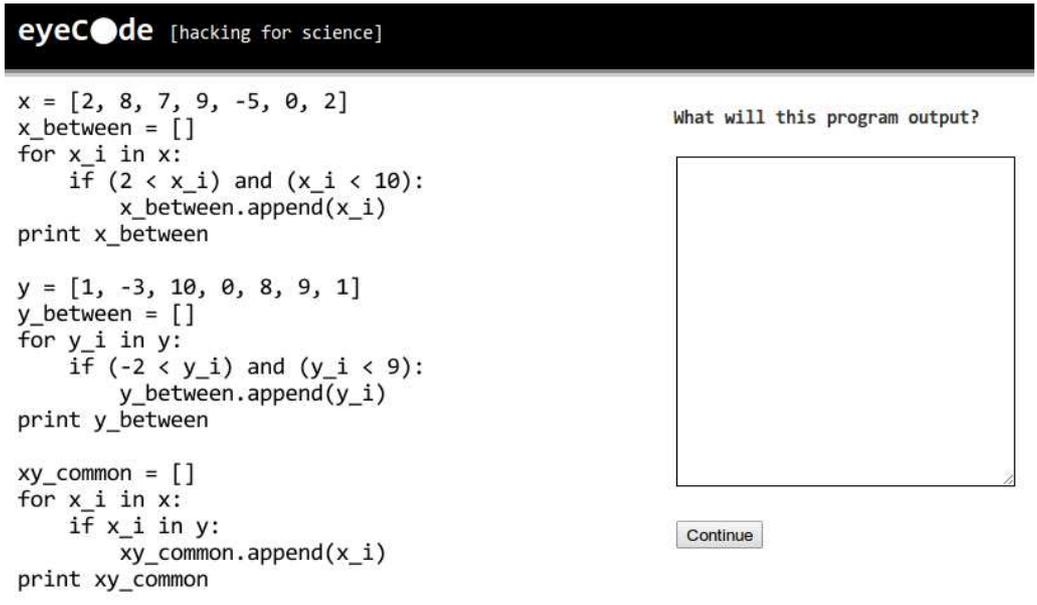}
  \end{center}
  \caption{Sample trial from the experiment (\texttt{between inline}).
  Participants were asked to predict the exact output of ten Python programs.}
  \label{fig:trial}
\end{figure}

The experiment consisted of a pre-test survey, ten trials (one program each),
and a post-test survey. Before the experiment began, participants were given
access to a small Python ``refresher,'' which listed the code and output of
several small programs. The pre-test survey gathered information about the
participant's age, gender, education, and experience. Participants were then
asked to predict the printed output of ten short Python programs, one version
randomly chosen from each of ten program types (Figure~\ref{fig:trial}). The
presentation order and names of the programs were randomized, and all answers
were final.  Although every program produced error-free output, participants
were not informed of this fact beforehand. The post-test survey gauged a
participant's confidence in their answers and the perceived difficulty of the
task.

We collected a total of \numTrials~trials from \numParticipants~participants
starting \expStartDate~and ending \expEndDate. Trials were graded
semi-automatically using a custom grading program. A grade of 10 points was
assigned for responses that exactly matched the program's output
(\numPerfectTrials~out of \numTrials~trials). A \emph{correct} grade of 7-9
points was given when responses had the right numbers or letters, but incorrect
formatting -- e.g., wrong whitespace, commas, brackets.  Common errors were
given partial credit from 2 to 4 points, depending on correct formatting. All
other responses were manually graded by two graders whose instructions were to
give fewer than 5 points for incorrect responses, and to take off points for
incorrect formatting (clear intermediary calculations or comments were ignored).
Graders' responses were strongly correlated ($r(598) = 0.90$), so individual
trial grades were averaged. Trial response times ranged from \minRT~to
\maxRT~seconds. Outliers beyond two standard deviations of the mean (in log
space) were discarded (\numOutliers~of \numTrials~trials).  Participants had a
total of 45 minutes to complete the entire experiment (10 trials + surveys), and
were required to give an answer to each question.

We had a total of twenty-five Python programs in ten different categories. These
programs were designed to be understandable to a wide audience, and therefore
did not touch on Python features outside of a first or second introductory
programming course. The programs ranged in size from 3 to 24 lines of code
(LOC).  Their cyclomatic complexities (CC) ranged from 1 to 7, and were
moderately correlated with LOC ($r(25) = 0.46$, $p < 0.05$). CC was computed
using the open source package PyMetrics. \\

\programname{Mechanical Turk} 
One hundred and thirty participants from Mechanical Turk completed the
experiment.  Workers were required to pass a Python pre-test, and could only
participate once. All code was displayed as an image, making it difficult to
copy/paste the code into a Python interpreter for quick answers.  All responses
were manually screened, and restarted trials or unfinished experiments were
discarded.

\section{Results}

Data analysis was performed in R, and all regressions were done with R's
built-in \texttt{lm} and \texttt{glm} functions. For linear regressions
involving the dependent measures grade and RT, we report intercepts and
coefficients ($\beta$). For logistic regressions (probability of a correct
answer or a common error), we report base predictor levels and the odds ratios
(OR).  Odds ratios can be difficult to interpret, and are often confused with
relative risk~\cite{davies1998can}. While the direction of an effect is usually
apparent, we caution against their interpretation as effect sizes (especially
when OR $< 1$).

Table~\ref{fig:results} summarizes the results in terms of average grades, mean
response times (RT), and significant effects (discussed in detail below).
Participants did well overall, and the plurality of trials resulted in perfect
responses (\numPerfectTrials~out of \numTrials). Years of Python experience was
a significant predictor of overall grade (intercept = $77.0$, $\beta = 1.23$, $p
< 0.05$), and a highly significant predictor of giving a perfect response (base
= $1.42$, OR = $1.09$, $p < 0.01$).  We discuss grade and RT differences between
program versions below.

\programname{between (2 versions)}
This program filters two lists of integers (\texttt{x} and \texttt{y}), prints
the results, and then prints the numbers that \texttt{x} and \texttt{y} have in
common. The \texttt{functions} version abstracts the \texttt{between} and
\texttt{common} operations into reusable functions, while the \texttt{inline}
version inlines the code, duplicating as necessary.

Because this is the longest and most complex program (in terms of CC), we
expected more experienced programmers to be faster and to make fewer errors. We
were surprised to find a significant effect of Python experience on the
probability of making a very specific error (base = $0.13$, OR = $1.44$, $p <
0.5$).  Instead of \texttt{[8, 9, 0]} for their last line of output, 22\% of
participants wrote \texttt{[8]} (for both versions). After the experiment, one
participant reported they mistakenly performed the ``common'' operation on lists
\texttt{x\_btwn} and \texttt{y\_btwn} instead of \texttt{x} and \texttt{y}
because it seemed like the next logical step. If others made the same mistake,
this may suggest an addition to Soloway's Rules of
Discourse~\cite{soloway1984empirical}: later computations should follow from
earlier ones. We hypothesize that moving the \texttt{common} operation code
before the two instances of \texttt{between} would eliminate this error.

\programname{counting (2 versions)}
This simple program loops through the range \texttt{[1, 2, 3, 4]}, printing
``The count is \texttt{i}'' and then ``Done counting'' for each number
\texttt{i}. The \textbf{nospace} version has the ``Done counting''
\texttt{print} statement immediately following ``The count is \texttt{i},''
whereas the \textbf{twospaces} version has two blank lines in between. Python is
sensitive to horizontal whitespace, but not vertical, so the extra lines do not
change the output of the program.

We expected more participants to mistakenly assume that the ``Done counting''
\texttt{print} statement was not part of the loop body in the \texttt{twospaces}
version.  This was the case: 59\% of responses in the \texttt{twospaces} version
contained this error as opposed to only 15\% in the \texttt{nospace} version
(ref = \texttt{nospace}, OR = $4.0$, $p < 0.0001$).  Blank lines, while not
usually syntactically relevant, are positively correlated with code
readability~\cite{buse2010learning}. We did not find a significant effect of
experience on the likelihood of making this mistake, suggesting that experts and
novices alike may benefit from an ending delimiter (e.g., an \texttt{end}
keyword or brackets).

\programname{funcall (3 versions)}
This program prints the product $f(1)*f(0)*f(-1)$ where $f(x) = x + 4$. The
\textbf{nospace} version has no spaces between the calls to \texttt{f}, while
the \textbf{space} version has a space before and after each \texttt{*} (e.g.,
$f(1)~*~f(0)~*~f(-1)$). The \textbf{vars} version saves the result of each call
to \texttt{f} in a separate variable, and then prints the product of these
variables. Code for \texttt{funcall}'s is not included for space reasons.

Most people were able to get the correct answer of 60 in 30 seconds or less. The
most common errors (only 7\% of responses) were 0, -60, and -80. We hypothesize
that these correspond to the following calculation errors: assuming $f(0) = 0$,
$f(-1) = -3$, and $f(-1) = -4$. There were no significant effects of version or
experience on grade.

\programname{initvar (3 versions)}
The \texttt{initvar} program computes the product and sum of variables
\texttt{a} and \texttt{b}. In the \texttt{good} version, \texttt{a} is
initialized to 1, so the product computes $4! = 24$, and \texttt{b} is
initialized to 0, making the summation $10$. In the \texttt{onebad} version, $b
= 1$, offsetting the summation by 1. In the \texttt{bothbad} version, $b = 1$
and $a = 0$, which makes the product 0.

We expected experienced programmers to make more errors due to the close
resemblance of code in the \texttt{*bad} versions to common operations performed
in the \texttt{good} version (factorial and summation). Instead, we found a
significant negative effect of the \texttt{good} version on grade (intercept =
$8.67$, $\beta = -1.52$, $p < 0.05$), which is likely due to the difficulty of
mentally performing 4 factorial. In the \texttt{bothbad} version, $a = 0$,
allowing participants to short-circuit the multiplication (since $a$ times
anything is still zero). The \texttt{onebad} version, which also required
performing the factorial, had a negative but non-significant effect on grade
($\beta = -0.97$).

\programname{order (2 versions)}
The \texttt{order} program prints the values of three functions, $f(x)$, $g(x)$,
and $h(x)$. In the \texttt{inorder} version, $f$, $g$, and $h$ are defined and
called in the same order. The \texttt{shuffled} version defines them out of
order ($h$, $f$, $g$).

We expected programmers to be slower when solving the \texttt{shuffled} version,
due to an implicit expectation that function definitions and use would follow
the same order. When including years of Python experience, we found a
significant main effect on RT of the \texttt{shuffled} version (intercept =
$54.3$, $\beta = 21.0$, $p < 0.05$) as well as an interaction between experience
and \texttt{shuffled} ($\beta = -7.1$, $p < 0.05$). Functions defined out of
order had a significant impact on response time, but experience helps
counter-act the effect.

\programname{overload (3 versions)}
This program uses the overloaded \texttt{+} operator, which serves as addition
for integers and concatenation for strings.  The \texttt{plusmixed} version uses
both overloads of the operator (\texttt{3 + 7}, \texttt{"5" + "3"}), while the
\texttt{multmixed} version and \texttt{strings} version only use \texttt{+} for
string concatenation (\texttt{"5" + "3"}).

We expected programmers in the \texttt{plusmixed} version to make the mistake of
interpreting \texttt{"5" + "3"} as 8 instead of \texttt{"53"} more often due to
the priming of \texttt{+} as addition instead of concatenation. While this error
occurred in about 11\% of responses across all versions, we did not see a
significant grade difference between versions. For response time, a significant
interaction between overall programming experience and the \texttt{plusmixed}
version was found (intercept = 42.5, $\beta = 3.34$, $p < 0.01$). Experienced
programmers were slowed down more by the \texttt{plusmixed} version than
inexperienced programmers, perhaps due to increased expectations that clustered
uses of \texttt{+} should correspond to the same operation (addition \textbf{or}
concatenation).

\programname{partition (3 versions)}
The \texttt{partition} program iterates through the ranges $[1,4]$ (unbalanced)
or $[1,5]$ (balanced), printing out \texttt{i low} for $i < 3$ and \texttt{i
high} for $i > 3$. The \texttt{balanced} version outputs two \texttt{low} and
two \texttt{high} lines, whereas the \texttt{unbalanced} versions produce two
\texttt{low} lines and only one \texttt{high} line. The
\texttt{unbalanced\_pivot} version calls attention to 3 by assigning it to a
variable named \texttt{pivot}.

We expected participants in the \texttt{unbalanced*} versions to add an
additional \texttt{high} line because there were four numbers in the list
(making it desirable for there to be four lines in the output). While there were
a handful of responses like this, the most common error was simply leaving off
the numbers on each line (e.g., \texttt{low} instead of \texttt{1 low}).
Programmers seeing the \texttt{unbalanced} version were less susceptible to this
error (ref = \texttt{balanced}, OR = $0.05$, $p < 0.05$), though we saw no
effect for the \texttt{unbalanced\_pivot} version. More programming experience
also helped participants avoid this kind of mistake across versions (base =
$1.66$, OR = $0.67$, $p < 0.05$). We hypothesize that the \texttt{balanced} and
\texttt{unbalanced\_pivot} versions matched a ``partition'' schema for
programmers, making them less likely to pay close attention to the loop body.

% 3 classified as high in 5/7 cases

\programname{rectangle (3 versions)}
This program computes the areas of two rectangles using an \texttt{area}
function with $x$ and $y$ scalar variables (\texttt{basic} version), $(x,y)$
coordinate pairs (\texttt{tuples} version), or a \texttt{Rectangle} class
(\texttt{class} version).

We expected participants seeing the \texttt{tuples} and \texttt{class} versions
to take longer, because these versions contain more complicated structures.
Almost everyone gave the correct answer, so there were no significant grade
differences between versions. We found a significant RT main effect for the
\texttt{tuples} version (intercept = $53.5$, $\beta = 60.4$, $p < 0.01$), and an
interaction between this version and Python experience ($\beta = -34.1$, $p <
0.01$).  Programmers in the \texttt{tuples} version took longer than those in
the \texttt{basic} version, but additional Python experience helped reverse this
effect. Surprisingly, we did not observe even a marginally significant RT effect
for the \texttt{classes} version, despite it being the longest program of the
three (21 lines vs. 14 and 18).

\programname{scope (2 versions)}
This program applies four functions to a variable named \texttt{added}: two
\texttt{add\_1} functions, and two \texttt{twice} functions. The
\texttt{samename} version reused the name \texttt{added} for function
parameters, while the \texttt{diffname} version used \texttt{num}. Because
Python uses ``pass by value'' semantics with integers, and because neither of
the functions return a value, \texttt{added} retains its initial value of 4
throughout the program (instead of being 22). This directly violates one of
Soloway's Rules of Discourse~\cite{soloway1984empirical}: do not include code
that will not be used.

We expected participants to mistakenly assume that the value of \texttt{added}
was changed more often when the parameter names of \texttt{add\_1} and
\texttt{twice} were both also named \texttt{added} (\texttt{samename} version).
There was marginally significant evidence for this ($p = 0.09$), but it was not
conclusive.  Additional Python experience helped reduce the likelihood of
answering 22 (base = $1.28$, OR = $0.71$, $p < 0.05$), but around half of the
participants still answered incorrectly.

\programname{whitespace (2 versions)}
The \texttt{whitespace} program prints the result of three simple linear
calculations. In the \texttt{zigzag} version, the code is laid out with one
space between every mathematical operation, so that the line endings have a
typical zig-zag appearance. The \texttt{linedup} version aligns each block of
code by its mathematical operators.

%TODO: Rewrite
We expected there to be a speed difference between the two versions in favor of
\texttt{linedup}. When designing the experiment, most of our pilot participants
agreed that this version was easier to read. The data did not support this
claim, but there was a significant effect on the probability of not respecting
order of operations. For the \texttt{zigzag} version, participants were
significantly more likely to incorrectly answer 5, 10, and 15 for the $y$ column
(ref = \texttt{linedup}, OR = $0.18$, $p < 0.05$). These are the answers that
would be obtained if a participant executed the multiplications before the
additions, contra the established of order of operations of Python and
mathematics more generally. This suggests that when computing the $y$ values,
participants in the \texttt{zigzag} version did addition before multiplication
more often than in \texttt{linedup} version. Effects of spacing on the perceived
order of arithmetic operations has been studied
before~\cite{goldstone2010education}, and our results indicate that spacing in
code layout also has an impact on order of executed operations.

\section{Discussion}

Experience helps experts in situations where they have reason to monitor for
specific kinds of errors, but may hurt in cases for which they have not been
trained. For example, our results from the \texttt{order} programs show that
experience protects programmers from being sensitive to the shuffled order of
the functions, because it is often the case in real world programs that
functions are defined and used out of order. However, experience leads to more
of a tendency to be primed in the \texttt{overload} programs because it is
unusual to use \texttt{+} for addition and then immediately for string
concatenation.  Real programs tend to have clumps of similar usage of an
operator, and programmers learn to be efficient by taking advantage of those
frequently occurring repetitions.  This same effect can be seen in
\texttt{between} programs, where experience leads to the expectation that the
\texttt{common} operation should immediately use the results of the
\texttt{between} operations.  Expectations are sometimes so strong, however,
that experience only plays a small role in avoiding errors. Programmers in both
versions of the \texttt{scope} program strongly expected the \texttt{add\_1} and
\texttt{twice} functions to do what their names implied, despite Python's
call-by-value semantics for integers and the fact that neither function actually
returned a value.

The physical aspects of notation, often considered superficial, can have a
profound impact on performance. Programmers were more likely to respect the
order of mathematical operations in the \texttt{linedup} version of
\texttt{whitespace}, showing how horizontal space can emphasize the common
structure between related calculations. Similarly, the \texttt{twospaces}
version of \texttt{counting} demonstrated that vertical space is more important
then indentation to programmers when judging whether or not statements belong to
the same loop body. Programmers often group blocks of related statements
together using vertical whitespace, but our results indicate that this seemingly
superficial space can cause even experienced programmers to internalize the
wrong program. Notation can also make a simple program more difficult to read.
Programmers took longer to respond to the \texttt{tuples} version of
\texttt{rectangle} despite it having fewer lines than the \texttt{class}
version. It is not uncommon in Python to use tuples for $(x, y)$ coordinates,
but the syntactic ``noise'' that is present in the \texttt{tuples} version for
variable names (e.g., \texttt{r1\_xy\_1}) and calculations (e.g., \texttt{width
= xy\_2[0] - xy\_1[0]}) likely gave programmers pause when verifying the code's
operation.

\programname{Future Work}
During the course of the experiment, Bloomington participants were seated in
front of a Tobii X300 eye-tracker. We plan to analyze this eye-tracking data,
and correlate it with our findings here. Specifically, we hope to see how code
features and experience effect the visual search process and, by proxy, program
comprehension.

\section{Acknowledgments}

Grant R305A1100060 from the Institute of Education Sciences Department of
Education and grant 0910218 from the National Science Foundation REESE supported
this research.

\bibliographystyle{plain}
\bibliography{local}

\newpage
\appendix
\section{Appendix}

\subsection{Programs}

\subsubsection{between - functions}
\begin{Verbatim}[commandchars=\\\{\},numbers=left,firstnumber=1,stepnumber=1]
\PY{k}{def} \PY{n+nf}{between}\PY{p}{(}\PY{n}{numbers}\PY{p}{,} \PY{n}{low}\PY{p}{,} \PY{n}{high}\PY{p}{)}\PY{p}{:}
    \PY{n}{winners} \PY{o}{=} \PY{p}{[}\PY{p}{]}
    \PY{k}{for} \PY{n}{num} \PY{o+ow}{in} \PY{n}{numbers}\PY{p}{:}
        \PY{k}{if} \PY{p}{(}\PY{n}{low} \PY{o}{\PYZlt{}} \PY{n}{num}\PY{p}{)} \PY{o+ow}{and} \PY{p}{(}\PY{n}{num} \PY{o}{\PYZlt{}} \PY{n}{high}\PY{p}{)}\PY{p}{:}
            \PY{n}{winners}\PY{o}{.}\PY{n}{append}\PY{p}{(}\PY{n}{num}\PY{p}{)}
    \PY{k}{return} \PY{n}{winners}

\PY{k}{def} \PY{n+nf}{common}\PY{p}{(}\PY{n}{list1}\PY{p}{,} \PY{n}{list2}\PY{p}{)}\PY{p}{:}
    \PY{n}{winners} \PY{o}{=} \PY{p}{[}\PY{p}{]}
    \PY{k}{for} \PY{n}{item1} \PY{o+ow}{in} \PY{n}{list1}\PY{p}{:}
        \PY{k}{if} \PY{n}{item1} \PY{o+ow}{in} \PY{n}{list2}\PY{p}{:}
            \PY{n}{winners}\PY{o}{.}\PY{n}{append}\PY{p}{(}\PY{n}{item1}\PY{p}{)}
    \PY{k}{return} \PY{n}{winners}

\PY{n}{x} \PY{o}{=} \PY{p}{[}\PY{l+m+mi}{2}\PY{p}{,} \PY{l+m+mi}{8}\PY{p}{,} \PY{l+m+mi}{7}\PY{p}{,} \PY{l+m+mi}{9}\PY{p}{,} \PY{o}{\PYZhy{}}\PY{l+m+mi}{5}\PY{p}{,} \PY{l+m+mi}{0}\PY{p}{,} \PY{l+m+mi}{2}\PY{p}{]}
\PY{n}{x\PYZus{}btwn} \PY{o}{=} \PY{n}{between}\PY{p}{(}\PY{n}{x}\PY{p}{,} \PY{l+m+mi}{2}\PY{p}{,} \PY{l+m+mi}{10}\PY{p}{)}
\PY{k}{print} \PY{n}{x\PYZus{}btwn} 

\PY{n}{y} \PY{o}{=} \PY{p}{[}\PY{l+m+mi}{1}\PY{p}{,} \PY{o}{\PYZhy{}}\PY{l+m+mi}{3}\PY{p}{,} \PY{l+m+mi}{10}\PY{p}{,} \PY{l+m+mi}{0}\PY{p}{,} \PY{l+m+mi}{8}\PY{p}{,} \PY{l+m+mi}{9}\PY{p}{,} \PY{l+m+mi}{1}\PY{p}{]}
\PY{n}{y\PYZus{}btwn} \PY{o}{=} \PY{n}{between}\PY{p}{(}\PY{n}{y}\PY{p}{,} \PY{o}{\PYZhy{}}\PY{l+m+mi}{2}\PY{p}{,} \PY{l+m+mi}{9}\PY{p}{)}
\PY{k}{print} \PY{n}{y\PYZus{}btwn} 

\PY{n}{xy\PYZus{}common} \PY{o}{=} \PY{n}{common}\PY{p}{(}\PY{n}{x}\PY{p}{,} \PY{n}{y}\PY{p}{)}
\PY{k}{print} \PY{n}{xy\PYZus{}common} 
\end{Verbatim}
\textbf{Output:}
\begin{lstlisting}
[8, 7, 9]
[1, 0, 8, 1]
[8, 9, 0]
\end{lstlisting}

\subsubsection{between - inline}
\begin{Verbatim}[commandchars=\\\{\},numbers=left,firstnumber=1,stepnumber=1]
\PY{n}{x} \PY{o}{=} \PY{p}{[}\PY{l+m+mi}{2}\PY{p}{,} \PY{l+m+mi}{8}\PY{p}{,} \PY{l+m+mi}{7}\PY{p}{,} \PY{l+m+mi}{9}\PY{p}{,} \PY{o}{\PYZhy{}}\PY{l+m+mi}{5}\PY{p}{,} \PY{l+m+mi}{0}\PY{p}{,} \PY{l+m+mi}{2}\PY{p}{]}
\PY{n}{x\PYZus{}between} \PY{o}{=} \PY{p}{[}\PY{p}{]}
\PY{k}{for} \PY{n}{x\PYZus{}i} \PY{o+ow}{in} \PY{n}{x}\PY{p}{:}
    \PY{k}{if} \PY{p}{(}\PY{l+m+mi}{2} \PY{o}{\PYZlt{}} \PY{n}{x\PYZus{}i}\PY{p}{)} \PY{o+ow}{and} \PY{p}{(}\PY{n}{x\PYZus{}i} \PY{o}{\PYZlt{}} \PY{l+m+mi}{10}\PY{p}{)}\PY{p}{:}
        \PY{n}{x\PYZus{}between}\PY{o}{.}\PY{n}{append}\PY{p}{(}\PY{n}{x\PYZus{}i}\PY{p}{)}
\PY{k}{print} \PY{n}{x\PYZus{}between}

\PY{n}{y} \PY{o}{=} \PY{p}{[}\PY{l+m+mi}{1}\PY{p}{,} \PY{o}{\PYZhy{}}\PY{l+m+mi}{3}\PY{p}{,} \PY{l+m+mi}{10}\PY{p}{,} \PY{l+m+mi}{0}\PY{p}{,} \PY{l+m+mi}{8}\PY{p}{,} \PY{l+m+mi}{9}\PY{p}{,} \PY{l+m+mi}{1}\PY{p}{]}
\PY{n}{y\PYZus{}between} \PY{o}{=} \PY{p}{[}\PY{p}{]}
\PY{k}{for} \PY{n}{y\PYZus{}i} \PY{o+ow}{in} \PY{n}{y}\PY{p}{:}
    \PY{k}{if} \PY{p}{(}\PY{o}{\PYZhy{}}\PY{l+m+mi}{2} \PY{o}{\PYZlt{}} \PY{n}{y\PYZus{}i}\PY{p}{)} \PY{o+ow}{and} \PY{p}{(}\PY{n}{y\PYZus{}i} \PY{o}{\PYZlt{}} \PY{l+m+mi}{9}\PY{p}{)}\PY{p}{:}
        \PY{n}{y\PYZus{}between}\PY{o}{.}\PY{n}{append}\PY{p}{(}\PY{n}{y\PYZus{}i}\PY{p}{)}
\PY{k}{print} \PY{n}{y\PYZus{}between}

\PY{n}{xy\PYZus{}common} \PY{o}{=} \PY{p}{[}\PY{p}{]}
\PY{k}{for} \PY{n}{x\PYZus{}i} \PY{o+ow}{in} \PY{n}{x}\PY{p}{:}
    \PY{k}{if} \PY{n}{x\PYZus{}i} \PY{o+ow}{in} \PY{n}{y}\PY{p}{:}
        \PY{n}{xy\PYZus{}common}\PY{o}{.}\PY{n}{append}\PY{p}{(}\PY{n}{x\PYZus{}i}\PY{p}{)}
\PY{k}{print} \PY{n}{xy\PYZus{}common}
\end{Verbatim}
\textbf{Output:}
\begin{lstlisting}
[8, 7, 9]
[1, 0, 8, 1]
[8, 9, 0]
\end{lstlisting}

\subsubsection{counting - nospace}
\begin{Verbatim}[commandchars=\\\{\},numbers=left,firstnumber=1,stepnumber=1]
\PY{k}{for} \PY{n}{i} \PY{o+ow}{in} \PY{p}{[}\PY{l+m+mi}{1}\PY{p}{,} \PY{l+m+mi}{2}\PY{p}{,} \PY{l+m+mi}{3}\PY{p}{,} \PY{l+m+mi}{4}\PY{p}{]}\PY{p}{:}
    \PY{k}{print} \PY{l+s}{\PYZdq{}}\PY{l+s}{The count is}\PY{l+s}{\PYZdq{}}\PY{p}{,} \PY{n}{i}
    \PY{k}{print} \PY{l+s}{\PYZdq{}}\PY{l+s}{Done counting}\PY{l+s}{\PYZdq{}}
\end{Verbatim}
\textbf{Output:}
\begin{lstlisting}
The count is 1
Done counting
The count is 2
Done counting
The count is 3
Done counting
The count is 4
Done counting
\end{lstlisting}

\subsubsection{counting - twospaces}
\begin{Verbatim}[commandchars=\\\{\},numbers=left,firstnumber=1,stepnumber=1]
\PY{k}{for} \PY{n}{i} \PY{o+ow}{in} \PY{p}{[}\PY{l+m+mi}{1}\PY{p}{,} \PY{l+m+mi}{2}\PY{p}{,} \PY{l+m+mi}{3}\PY{p}{,} \PY{l+m+mi}{4}\PY{p}{]}\PY{p}{:}
    \PY{k}{print} \PY{l+s}{\PYZdq{}}\PY{l+s}{The count is}\PY{l+s}{\PYZdq{}}\PY{p}{,} \PY{n}{i} 


    \PY{k}{print} \PY{l+s}{\PYZdq{}}\PY{l+s}{Done counting}\PY{l+s}{\PYZdq{}} 
\end{Verbatim}
\textbf{Output:}
\begin{lstlisting}
The count is 1
Done counting
The count is 2
Done counting
The count is 3
Done counting
The count is 4
Done counting
\end{lstlisting}

\subsubsection{funcall - nospace}
\begin{Verbatim}[commandchars=\\\{\},numbers=left,firstnumber=1,stepnumber=1]
\PY{k}{def} \PY{n+nf}{f}\PY{p}{(}\PY{n}{x}\PY{p}{)}\PY{p}{:}
    \PY{k}{return} \PY{n}{x} \PY{o}{+} \PY{l+m+mi}{4}

\PY{k}{print} \PY{n}{f}\PY{p}{(}\PY{l+m+mi}{1}\PY{p}{)}\PY{o}{*}\PY{n}{f}\PY{p}{(}\PY{l+m+mi}{0}\PY{p}{)}\PY{o}{*}\PY{n}{f}\PY{p}{(}\PY{o}{\PYZhy{}}\PY{l+m+mi}{1}\PY{p}{)}
\end{Verbatim}
\textbf{Output:}
\begin{lstlisting}
60
\end{lstlisting}

\subsubsection{funcall - space}
\begin{Verbatim}[commandchars=\\\{\},numbers=left,firstnumber=1,stepnumber=1]
\PY{k}{def} \PY{n+nf}{f}\PY{p}{(}\PY{n}{x}\PY{p}{)}\PY{p}{:}
    \PY{k}{return} \PY{n}{x} \PY{o}{+} \PY{l+m+mi}{4}

\PY{k}{print} \PY{n}{f}\PY{p}{(}\PY{l+m+mi}{1}\PY{p}{)} \PY{o}{*} \PY{n}{f}\PY{p}{(}\PY{l+m+mi}{0}\PY{p}{)} \PY{o}{*} \PY{n}{f}\PY{p}{(}\PY{o}{\PYZhy{}}\PY{l+m+mi}{1}\PY{p}{)}
\end{Verbatim}
\textbf{Output:}
\begin{lstlisting}
60
\end{lstlisting}

\subsubsection{funcall - vars}
\begin{Verbatim}[commandchars=\\\{\},numbers=left,firstnumber=1,stepnumber=1]
\PY{k}{def} \PY{n+nf}{f}\PY{p}{(}\PY{n}{x}\PY{p}{)}\PY{p}{:}
    \PY{k}{return} \PY{n}{x} \PY{o}{+} \PY{l+m+mi}{4}

\PY{n}{x} \PY{o}{=} \PY{n}{f}\PY{p}{(}\PY{l+m+mi}{1}\PY{p}{)}
\PY{n}{y} \PY{o}{=} \PY{n}{f}\PY{p}{(}\PY{l+m+mi}{0}\PY{p}{)}
\PY{n}{z} \PY{o}{=} \PY{n}{f}\PY{p}{(}\PY{o}{\PYZhy{}}\PY{l+m+mi}{1}\PY{p}{)}
\PY{k}{print} \PY{n}{x} \PY{o}{*} \PY{n}{y} \PY{o}{*} \PY{n}{z}
\end{Verbatim}
\textbf{Output:}
\begin{lstlisting}
60
\end{lstlisting}

\subsubsection{initvar - bothbad}
\begin{Verbatim}[commandchars=\\\{\},numbers=left,firstnumber=1,stepnumber=1]
\PY{n}{a} \PY{o}{=} \PY{l+m+mi}{0}
\PY{k}{for} \PY{n}{i} \PY{o+ow}{in} \PY{p}{[}\PY{l+m+mi}{1}\PY{p}{,} \PY{l+m+mi}{2}\PY{p}{,} \PY{l+m+mi}{3}\PY{p}{,} \PY{l+m+mi}{4}\PY{p}{]}\PY{p}{:}
    \PY{n}{a} \PY{o}{=} \PY{n}{a} \PY{o}{*} \PY{n}{i}
\PY{k}{print} \PY{n}{a}

\PY{n}{b} \PY{o}{=} \PY{l+m+mi}{1}
\PY{k}{for} \PY{n}{i} \PY{o+ow}{in} \PY{p}{[}\PY{l+m+mi}{1}\PY{p}{,} \PY{l+m+mi}{2}\PY{p}{,} \PY{l+m+mi}{3}\PY{p}{,} \PY{l+m+mi}{4}\PY{p}{]}\PY{p}{:}
    \PY{n}{b} \PY{o}{=} \PY{n}{b} \PY{o}{+} \PY{n}{i}
\PY{k}{print} \PY{n}{b}
\end{Verbatim}
\textbf{Output:}
\begin{lstlisting}
0
11
\end{lstlisting}

\subsubsection{initvar - good}
\begin{Verbatim}[commandchars=\\\{\},numbers=left,firstnumber=1,stepnumber=1]
\PY{n}{a} \PY{o}{=} \PY{l+m+mi}{1}
\PY{k}{for} \PY{n}{i} \PY{o+ow}{in} \PY{p}{[}\PY{l+m+mi}{1}\PY{p}{,} \PY{l+m+mi}{2}\PY{p}{,} \PY{l+m+mi}{3}\PY{p}{,} \PY{l+m+mi}{4}\PY{p}{]}\PY{p}{:}
    \PY{n}{a} \PY{o}{=} \PY{n}{a} \PY{o}{*} \PY{n}{i}
\PY{k}{print} \PY{n}{a}

\PY{n}{b} \PY{o}{=} \PY{l+m+mi}{0}
\PY{k}{for} \PY{n}{i} \PY{o+ow}{in} \PY{p}{[}\PY{l+m+mi}{1}\PY{p}{,} \PY{l+m+mi}{2}\PY{p}{,} \PY{l+m+mi}{3}\PY{p}{,} \PY{l+m+mi}{4}\PY{p}{]}\PY{p}{:}
    \PY{n}{b} \PY{o}{=} \PY{n}{b} \PY{o}{+} \PY{n}{i}
\PY{k}{print} \PY{n}{b}
\end{Verbatim}
\textbf{Output:}
\begin{lstlisting}
24
10
\end{lstlisting}

\subsubsection{initvar - onebad}
\begin{Verbatim}[commandchars=\\\{\},numbers=left,firstnumber=1,stepnumber=1]
\PY{n}{a} \PY{o}{=} \PY{l+m+mi}{1}
\PY{k}{for} \PY{n}{i} \PY{o+ow}{in} \PY{p}{[}\PY{l+m+mi}{1}\PY{p}{,} \PY{l+m+mi}{2}\PY{p}{,} \PY{l+m+mi}{3}\PY{p}{,} \PY{l+m+mi}{4}\PY{p}{]}\PY{p}{:}
    \PY{n}{a} \PY{o}{=} \PY{n}{a} \PY{o}{*} \PY{n}{i}
\PY{k}{print} \PY{n}{a}

\PY{n}{b} \PY{o}{=} \PY{l+m+mi}{1}
\PY{k}{for} \PY{n}{i} \PY{o+ow}{in} \PY{p}{[}\PY{l+m+mi}{1}\PY{p}{,} \PY{l+m+mi}{2}\PY{p}{,} \PY{l+m+mi}{3}\PY{p}{,} \PY{l+m+mi}{4}\PY{p}{]}\PY{p}{:}
    \PY{n}{b} \PY{o}{=} \PY{n}{b} \PY{o}{+} \PY{n}{i}
\PY{k}{print} \PY{n}{b}
\end{Verbatim}
\textbf{Output:}
\begin{lstlisting}
24
11
\end{lstlisting}

\subsubsection{order - inorder}
\begin{Verbatim}[commandchars=\\\{\},numbers=left,firstnumber=1,stepnumber=1]
\PY{k}{def} \PY{n+nf}{f}\PY{p}{(}\PY{n}{x}\PY{p}{)}\PY{p}{:}
    \PY{k}{return} \PY{n}{x} \PY{o}{+} \PY{l+m+mi}{4}

\PY{k}{def} \PY{n+nf}{g}\PY{p}{(}\PY{n}{x}\PY{p}{)}\PY{p}{:}
    \PY{k}{return} \PY{n}{x} \PY{o}{*} \PY{l+m+mi}{2}

\PY{k}{def} \PY{n+nf}{h}\PY{p}{(}\PY{n}{x}\PY{p}{)}\PY{p}{:}
    \PY{k}{return} \PY{n}{f}\PY{p}{(}\PY{n}{x}\PY{p}{)} \PY{o}{+} \PY{n}{g}\PY{p}{(}\PY{n}{x}\PY{p}{)}

\PY{n}{x} \PY{o}{=} \PY{l+m+mi}{1}
\PY{n}{a} \PY{o}{=} \PY{n}{f}\PY{p}{(}\PY{n}{x}\PY{p}{)}
\PY{n}{b} \PY{o}{=} \PY{n}{g}\PY{p}{(}\PY{n}{x}\PY{p}{)}
\PY{n}{c} \PY{o}{=} \PY{n}{h}\PY{p}{(}\PY{n}{x}\PY{p}{)}
\PY{k}{print} \PY{n}{a}\PY{p}{,} \PY{n}{b}\PY{p}{,} \PY{n}{c}
\end{Verbatim}
\textbf{Output:}
\begin{lstlisting}
5 2 7
\end{lstlisting}

\subsubsection{order - shuffled}
\begin{Verbatim}[commandchars=\\\{\},numbers=left,firstnumber=1,stepnumber=1]
\PY{k}{def} \PY{n+nf}{h}\PY{p}{(}\PY{n}{x}\PY{p}{)}\PY{p}{:}
    \PY{k}{return} \PY{n}{f}\PY{p}{(}\PY{n}{x}\PY{p}{)} \PY{o}{+} \PY{n}{g}\PY{p}{(}\PY{n}{x}\PY{p}{)}

\PY{k}{def} \PY{n+nf}{f}\PY{p}{(}\PY{n}{x}\PY{p}{)}\PY{p}{:}
    \PY{k}{return} \PY{n}{x} \PY{o}{+} \PY{l+m+mi}{4}

\PY{k}{def} \PY{n+nf}{g}\PY{p}{(}\PY{n}{x}\PY{p}{)}\PY{p}{:}
    \PY{k}{return} \PY{n}{x} \PY{o}{*} \PY{l+m+mi}{2}

\PY{n}{x} \PY{o}{=} \PY{l+m+mi}{1}
\PY{n}{a} \PY{o}{=} \PY{n}{f}\PY{p}{(}\PY{n}{x}\PY{p}{)}
\PY{n}{b} \PY{o}{=} \PY{n}{g}\PY{p}{(}\PY{n}{x}\PY{p}{)}
\PY{n}{c} \PY{o}{=} \PY{n}{h}\PY{p}{(}\PY{n}{x}\PY{p}{)}
\PY{k}{print} \PY{n}{a}\PY{p}{,} \PY{n}{b}\PY{p}{,} \PY{n}{c}
\end{Verbatim}
\textbf{Output:}
\begin{lstlisting}
5 2 7
\end{lstlisting}

\subsubsection{overload - multmixed}
\begin{Verbatim}[commandchars=\\\{\},numbers=left,firstnumber=1,stepnumber=1]
\PY{n}{a} \PY{o}{=} \PY{l+m+mi}{4}
\PY{n}{b} \PY{o}{=} \PY{l+m+mi}{3}
\PY{k}{print} \PY{n}{a} \PY{o}{*} \PY{n}{b}

\PY{n}{c} \PY{o}{=} \PY{l+m+mi}{7}
\PY{n}{d} \PY{o}{=} \PY{l+m+mi}{2}
\PY{k}{print} \PY{n}{c} \PY{o}{*} \PY{n}{d}

\PY{n}{e} \PY{o}{=} \PY{l+s}{\PYZdq{}}\PY{l+s}{5}\PY{l+s}{\PYZdq{}}
\PY{n}{f} \PY{o}{=} \PY{l+s}{\PYZdq{}}\PY{l+s}{3}\PY{l+s}{\PYZdq{}}
\PY{k}{print} \PY{n}{e} \PY{o}{+} \PY{n}{f}
\end{Verbatim}
\textbf{Output:}
\begin{lstlisting}
12
14
53
\end{lstlisting}

\subsubsection{overload - plusmixed}
\begin{Verbatim}[commandchars=\\\{\},numbers=left,firstnumber=1,stepnumber=1]
\PY{n}{a} \PY{o}{=} \PY{l+m+mi}{4}
\PY{n}{b} \PY{o}{=} \PY{l+m+mi}{3}
\PY{k}{print} \PY{n}{a} \PY{o}{+} \PY{n}{b}

\PY{n}{c} \PY{o}{=} \PY{l+m+mi}{7}
\PY{n}{d} \PY{o}{=} \PY{l+m+mi}{2}
\PY{k}{print} \PY{n}{c} \PY{o}{+} \PY{n}{d}

\PY{n}{e} \PY{o}{=} \PY{l+s}{\PYZdq{}}\PY{l+s}{5}\PY{l+s}{\PYZdq{}}
\PY{n}{f} \PY{o}{=} \PY{l+s}{\PYZdq{}}\PY{l+s}{3}\PY{l+s}{\PYZdq{}}
\PY{k}{print} \PY{n}{e} \PY{o}{+} \PY{n}{f}
\end{Verbatim}
\textbf{Output:}
\begin{lstlisting}
7
9
53
\end{lstlisting}

\subsubsection{overload - strings}
\begin{Verbatim}[commandchars=\\\{\},numbers=left,firstnumber=1,stepnumber=1]
\PY{n}{a} \PY{o}{=} \PY{l+s}{\PYZdq{}}\PY{l+s}{hi}\PY{l+s}{\PYZdq{}}
\PY{n}{b} \PY{o}{=} \PY{l+s}{\PYZdq{}}\PY{l+s}{bye}\PY{l+s}{\PYZdq{}}
\PY{k}{print} \PY{n}{a} \PY{o}{+} \PY{n}{b}

\PY{n}{c} \PY{o}{=} \PY{l+s}{\PYZdq{}}\PY{l+s}{street}\PY{l+s}{\PYZdq{}}
\PY{n}{d} \PY{o}{=} \PY{l+s}{\PYZdq{}}\PY{l+s}{penny}\PY{l+s}{\PYZdq{}}
\PY{k}{print} \PY{n}{c} \PY{o}{+} \PY{n}{d}

\PY{n}{e} \PY{o}{=} \PY{l+s}{\PYZdq{}}\PY{l+s}{5}\PY{l+s}{\PYZdq{}}
\PY{n}{f} \PY{o}{=} \PY{l+s}{\PYZdq{}}\PY{l+s}{3}\PY{l+s}{\PYZdq{}}
\PY{k}{print} \PY{n}{e} \PY{o}{+} \PY{n}{f}
\end{Verbatim}
\textbf{Output:}
\begin{lstlisting}
hibye
streetpenny
53
\end{lstlisting}

\subsubsection{partition - balanced}
\begin{Verbatim}[commandchars=\\\{\},numbers=left,firstnumber=1,stepnumber=1]
\PY{k}{for} \PY{n}{i} \PY{o+ow}{in} \PY{p}{[}\PY{l+m+mi}{1}\PY{p}{,} \PY{l+m+mi}{2}\PY{p}{,} \PY{l+m+mi}{3}\PY{p}{,} \PY{l+m+mi}{4}\PY{p}{,} \PY{l+m+mi}{5}\PY{p}{]}\PY{p}{:}
    \PY{k}{if} \PY{p}{(}\PY{n}{i} \PY{o}{\PYZlt{}} \PY{l+m+mi}{3}\PY{p}{)}\PY{p}{:}
        \PY{k}{print} \PY{n}{i}\PY{p}{,} \PY{l+s}{\PYZdq{}}\PY{l+s}{low}\PY{l+s}{\PYZdq{}}
    \PY{k}{if} \PY{p}{(}\PY{n}{i} \PY{o}{\PYZgt{}} \PY{l+m+mi}{3}\PY{p}{)}\PY{p}{:}
        \PY{k}{print} \PY{n}{i}\PY{p}{,} \PY{l+s}{\PYZdq{}}\PY{l+s}{high}\PY{l+s}{\PYZdq{}}
\end{Verbatim}
\textbf{Output:}
\begin{lstlisting}
1 low
2 low
4 high
5 high
\end{lstlisting}

\subsubsection{partition - unbalanced}
\begin{Verbatim}[commandchars=\\\{\},numbers=left,firstnumber=1,stepnumber=1]
\PY{k}{for} \PY{n}{i} \PY{o+ow}{in} \PY{p}{[}\PY{l+m+mi}{1}\PY{p}{,} \PY{l+m+mi}{2}\PY{p}{,} \PY{l+m+mi}{3}\PY{p}{,} \PY{l+m+mi}{4}\PY{p}{]}\PY{p}{:}
    \PY{k}{if} \PY{p}{(}\PY{n}{i} \PY{o}{\PYZlt{}} \PY{l+m+mi}{3}\PY{p}{)}\PY{p}{:}
        \PY{k}{print} \PY{n}{i}\PY{p}{,} \PY{l+s}{\PYZdq{}}\PY{l+s}{low}\PY{l+s}{\PYZdq{}}
    \PY{k}{if} \PY{p}{(}\PY{n}{i} \PY{o}{\PYZgt{}} \PY{l+m+mi}{3}\PY{p}{)}\PY{p}{:}
        \PY{k}{print} \PY{n}{i}\PY{p}{,} \PY{l+s}{\PYZdq{}}\PY{l+s}{high}\PY{l+s}{\PYZdq{}}
\end{Verbatim}
\textbf{Output:}
\begin{lstlisting}
1 low
2 low
4 high
\end{lstlisting}

\subsubsection{partition - unbalanced\_pivot}
\begin{Verbatim}[commandchars=\\\{\},numbers=left,firstnumber=1,stepnumber=1]
\PY{n}{pivot} \PY{o}{=} \PY{l+m+mi}{3}
\PY{k}{for} \PY{n}{i} \PY{o+ow}{in} \PY{p}{[}\PY{l+m+mi}{1}\PY{p}{,} \PY{l+m+mi}{2}\PY{p}{,} \PY{l+m+mi}{3}\PY{p}{,} \PY{l+m+mi}{4}\PY{p}{]}\PY{p}{:}
    \PY{k}{if} \PY{p}{(}\PY{n}{i} \PY{o}{\PYZlt{}} \PY{n}{pivot}\PY{p}{)}\PY{p}{:}
        \PY{k}{print} \PY{n}{i}\PY{p}{,} \PY{l+s}{\PYZdq{}}\PY{l+s}{low}\PY{l+s}{\PYZdq{}}
    \PY{k}{if} \PY{p}{(}\PY{n}{i} \PY{o}{\PYZgt{}} \PY{n}{pivot}\PY{p}{)}\PY{p}{:}
        \PY{k}{print} \PY{n}{i}\PY{p}{,} \PY{l+s}{\PYZdq{}}\PY{l+s}{high}\PY{l+s}{\PYZdq{}}
\end{Verbatim}
\textbf{Output:}
\begin{lstlisting}
1 low
2 low
4 high
\end{lstlisting}

\subsubsection{rectangle - basic}
\begin{Verbatim}[commandchars=\\\{\},numbers=left,firstnumber=1,stepnumber=1]
\PY{k}{def} \PY{n+nf}{area}\PY{p}{(}\PY{n}{x1}\PY{p}{,} \PY{n}{y1}\PY{p}{,} \PY{n}{x2}\PY{p}{,} \PY{n}{y2}\PY{p}{)}\PY{p}{:}
    \PY{n}{width} \PY{o}{=} \PY{n}{x2} \PY{o}{\PYZhy{}} \PY{n}{x1}
    \PY{n}{height} \PY{o}{=} \PY{n}{y2} \PY{o}{\PYZhy{}} \PY{n}{y1}
    \PY{k}{return} \PY{n}{width} \PY{o}{*} \PY{n}{height}

\PY{n}{r1\PYZus{}x1} \PY{o}{=} \PY{l+m+mi}{0}
\PY{n}{r1\PYZus{}y1} \PY{o}{=} \PY{l+m+mi}{0}
\PY{n}{r1\PYZus{}x2} \PY{o}{=} \PY{l+m+mi}{10}
\PY{n}{r1\PYZus{}y2} \PY{o}{=} \PY{l+m+mi}{10}
\PY{n}{r1\PYZus{}area} \PY{o}{=} \PY{n}{area}\PY{p}{(}\PY{n}{r1\PYZus{}x1}\PY{p}{,} \PY{n}{r1\PYZus{}y1}\PY{p}{,} \PY{n}{r1\PYZus{}x2}\PY{p}{,} \PY{n}{r1\PYZus{}y2}\PY{p}{)}
\PY{k}{print} \PY{n}{r1\PYZus{}area}

\PY{n}{r2\PYZus{}x1} \PY{o}{=} \PY{l+m+mi}{5}
\PY{n}{r2\PYZus{}y1} \PY{o}{=} \PY{l+m+mi}{5}
\PY{n}{r2\PYZus{}x2} \PY{o}{=} \PY{l+m+mi}{10}
\PY{n}{r2\PYZus{}y2} \PY{o}{=} \PY{l+m+mi}{10}
\PY{n}{r2\PYZus{}area} \PY{o}{=} \PY{n}{area}\PY{p}{(}\PY{n}{r2\PYZus{}x1}\PY{p}{,} \PY{n}{r2\PYZus{}y1}\PY{p}{,} \PY{n}{r2\PYZus{}x2}\PY{p}{,} \PY{n}{r2\PYZus{}y2}\PY{p}{)}
\PY{k}{print} \PY{n}{r2\PYZus{}area}
\end{Verbatim}
\textbf{Output:}
\begin{lstlisting}
100
25
\end{lstlisting}

\subsubsection{rectangle - class}
\begin{Verbatim}[commandchars=\\\{\},numbers=left,firstnumber=1,stepnumber=1]
\PY{k}{class} \PY{n+nc}{Rectangle}\PY{p}{:}
    \PY{k}{def} \PY{n+nf}{\PYZus{}\PYZus{}init\PYZus{}\PYZus{}}\PY{p}{(}\PY{n+nb+bp}{self}\PY{p}{,} \PY{n}{x1}\PY{p}{,} \PY{n}{y1}\PY{p}{,} \PY{n}{x2}\PY{p}{,} \PY{n}{y2}\PY{p}{)}\PY{p}{:}
        \PY{n+nb+bp}{self}\PY{o}{.}\PY{n}{x1} \PY{o}{=} \PY{n}{x1}
        \PY{n+nb+bp}{self}\PY{o}{.}\PY{n}{y1} \PY{o}{=} \PY{n}{y1}
        \PY{n+nb+bp}{self}\PY{o}{.}\PY{n}{x2} \PY{o}{=} \PY{n}{x2}
        \PY{n+nb+bp}{self}\PY{o}{.}\PY{n}{y2} \PY{o}{=} \PY{n}{y2}

    \PY{k}{def} \PY{n+nf}{width}\PY{p}{(}\PY{n+nb+bp}{self}\PY{p}{)}\PY{p}{:}
        \PY{k}{return} \PY{n+nb+bp}{self}\PY{o}{.}\PY{n}{x2} \PY{o}{\PYZhy{}} \PY{n+nb+bp}{self}\PY{o}{.}\PY{n}{x1}

    \PY{k}{def} \PY{n+nf}{height}\PY{p}{(}\PY{n+nb+bp}{self}\PY{p}{)}\PY{p}{:}
        \PY{k}{return} \PY{n+nb+bp}{self}\PY{o}{.}\PY{n}{y2} \PY{o}{\PYZhy{}} \PY{n+nb+bp}{self}\PY{o}{.}\PY{n}{y1}

    \PY{k}{def} \PY{n+nf}{area}\PY{p}{(}\PY{n+nb+bp}{self}\PY{p}{)}\PY{p}{:}
        \PY{k}{return} \PY{n+nb+bp}{self}\PY{o}{.}\PY{n}{width}\PY{p}{(}\PY{p}{)} \PY{o}{*} \PY{n+nb+bp}{self}\PY{o}{.}\PY{n}{height}\PY{p}{(}\PY{p}{)}

\PY{n}{rect1} \PY{o}{=} \PY{n}{Rectangle}\PY{p}{(}\PY{l+m+mi}{0}\PY{p}{,} \PY{l+m+mi}{0}\PY{p}{,} \PY{l+m+mi}{10}\PY{p}{,} \PY{l+m+mi}{10}\PY{p}{)}
\PY{k}{print} \PY{n}{rect1}\PY{o}{.}\PY{n}{area}\PY{p}{(}\PY{p}{)}

\PY{n}{rect2} \PY{o}{=} \PY{n}{Rectangle}\PY{p}{(}\PY{l+m+mi}{5}\PY{p}{,} \PY{l+m+mi}{5}\PY{p}{,} \PY{l+m+mi}{10}\PY{p}{,} \PY{l+m+mi}{10}\PY{p}{)}
\PY{k}{print} \PY{n}{rect2}\PY{o}{.}\PY{n}{area}\PY{p}{(}\PY{p}{)}
\end{Verbatim}
\textbf{Output:}
\begin{lstlisting}
100
25
\end{lstlisting}

\subsubsection{rectangle - tuples}
\begin{Verbatim}[commandchars=\\\{\},numbers=left,firstnumber=1,stepnumber=1]
\PY{k}{def} \PY{n+nf}{area}\PY{p}{(}\PY{n}{xy\PYZus{}1}\PY{p}{,} \PY{n}{xy\PYZus{}2}\PY{p}{)}\PY{p}{:}
    \PY{n}{width} \PY{o}{=} \PY{n}{xy\PYZus{}2}\PY{p}{[}\PY{l+m+mi}{0}\PY{p}{]} \PY{o}{\PYZhy{}} \PY{n}{xy\PYZus{}1}\PY{p}{[}\PY{l+m+mi}{0}\PY{p}{]}
    \PY{n}{height} \PY{o}{=} \PY{n}{xy\PYZus{}2}\PY{p}{[}\PY{l+m+mi}{1}\PY{p}{]} \PY{o}{\PYZhy{}} \PY{n}{xy\PYZus{}1}\PY{p}{[}\PY{l+m+mi}{1}\PY{p}{]}
    \PY{k}{return} \PY{n}{width} \PY{o}{*} \PY{n}{height}

\PY{n}{r1\PYZus{}xy\PYZus{}1} \PY{o}{=} \PY{p}{(}\PY{l+m+mi}{0}\PY{p}{,} \PY{l+m+mi}{0}\PY{p}{)}
\PY{n}{r1\PYZus{}xy\PYZus{}2} \PY{o}{=} \PY{p}{(}\PY{l+m+mi}{10}\PY{p}{,} \PY{l+m+mi}{10}\PY{p}{)}
\PY{n}{r1\PYZus{}area} \PY{o}{=} \PY{n}{area}\PY{p}{(}\PY{n}{r1\PYZus{}xy\PYZus{}1}\PY{p}{,} \PY{n}{r1\PYZus{}xy\PYZus{}2}\PY{p}{)}
\PY{k}{print} \PY{n}{r1\PYZus{}area}

\PY{n}{r2\PYZus{}xy\PYZus{}1} \PY{o}{=} \PY{p}{(}\PY{l+m+mi}{5}\PY{p}{,} \PY{l+m+mi}{5}\PY{p}{)}
\PY{n}{r2\PYZus{}xy\PYZus{}2} \PY{o}{=} \PY{p}{(}\PY{l+m+mi}{10}\PY{p}{,} \PY{l+m+mi}{10}\PY{p}{)}
\PY{n}{r2\PYZus{}area} \PY{o}{=} \PY{n}{area}\PY{p}{(}\PY{n}{r2\PYZus{}xy\PYZus{}1}\PY{p}{,} \PY{n}{r2\PYZus{}xy\PYZus{}2}\PY{p}{)}
\PY{k}{print} \PY{n}{r2\PYZus{}area}
\end{Verbatim}
\textbf{Output:}
\begin{lstlisting}
100
25
\end{lstlisting}

\subsubsection{scope - diffname}
\begin{Verbatim}[commandchars=\\\{\},numbers=left,firstnumber=1,stepnumber=1]
\PY{k}{def} \PY{n+nf}{add\PYZus{}1}\PY{p}{(}\PY{n}{num}\PY{p}{)}\PY{p}{:}
    \PY{n}{num} \PY{o}{=} \PY{n}{num} \PY{o}{+} \PY{l+m+mi}{1}

\PY{k}{def} \PY{n+nf}{twice}\PY{p}{(}\PY{n}{num}\PY{p}{)}\PY{p}{:}
    \PY{n}{num} \PY{o}{=} \PY{n}{num} \PY{o}{*} \PY{l+m+mi}{2}

\PY{n}{added} \PY{o}{=} \PY{l+m+mi}{4}
\PY{n}{add\PYZus{}1}\PY{p}{(}\PY{n}{added}\PY{p}{)}
\PY{n}{twice}\PY{p}{(}\PY{n}{added}\PY{p}{)}
\PY{n}{add\PYZus{}1}\PY{p}{(}\PY{n}{added}\PY{p}{)}
\PY{n}{twice}\PY{p}{(}\PY{n}{added}\PY{p}{)}
\PY{k}{print} \PY{n}{added}
\end{Verbatim}
\textbf{Output:}
\begin{lstlisting}
4
\end{lstlisting}

\subsubsection{scope - samename}
\begin{Verbatim}[commandchars=\\\{\},numbers=left,firstnumber=1,stepnumber=1]
\PY{k}{def} \PY{n+nf}{add\PYZus{}1}\PY{p}{(}\PY{n}{added}\PY{p}{)}\PY{p}{:}
    \PY{n}{added} \PY{o}{=} \PY{n}{added} \PY{o}{+} \PY{l+m+mi}{1}

\PY{k}{def} \PY{n+nf}{twice}\PY{p}{(}\PY{n}{added}\PY{p}{)}\PY{p}{:}
    \PY{n}{added} \PY{o}{=} \PY{n}{added} \PY{o}{*} \PY{l+m+mi}{2}

\PY{n}{added} \PY{o}{=} \PY{l+m+mi}{4}
\PY{n}{add\PYZus{}1}\PY{p}{(}\PY{n}{added}\PY{p}{)}
\PY{n}{twice}\PY{p}{(}\PY{n}{added}\PY{p}{)}
\PY{n}{add\PYZus{}1}\PY{p}{(}\PY{n}{added}\PY{p}{)}
\PY{n}{twice}\PY{p}{(}\PY{n}{added}\PY{p}{)}
\PY{k}{print} \PY{n}{added}
\end{Verbatim}
\textbf{Output:}
\begin{lstlisting}
4
\end{lstlisting}

\subsubsection{whitespace - linedup}
\begin{Verbatim}[commandchars=\\\{\},numbers=left,firstnumber=1,stepnumber=1]
\PY{n}{intercept} \PY{o}{=} \PY{l+m+mi}{1}
\PY{n}{slope}     \PY{o}{=} \PY{l+m+mi}{5}

\PY{n}{x\PYZus{}base}  \PY{o}{=} \PY{l+m+mi}{0}
\PY{n}{x\PYZus{}other} \PY{o}{=} \PY{n}{x\PYZus{}base} \PY{o}{+} \PY{l+m+mi}{1}
\PY{n}{x\PYZus{}end}   \PY{o}{=} \PY{n}{x\PYZus{}base} \PY{o}{+} \PY{n}{x\PYZus{}other} \PY{o}{+} \PY{l+m+mi}{1}

\PY{n}{y\PYZus{}base}  \PY{o}{=} \PY{n}{slope} \PY{o}{*} \PY{n}{x\PYZus{}base}  \PY{o}{+} \PY{n}{intercept}
\PY{n}{y\PYZus{}other} \PY{o}{=} \PY{n}{slope} \PY{o}{*} \PY{n}{x\PYZus{}other} \PY{o}{+} \PY{n}{intercept}
\PY{n}{y\PYZus{}end}   \PY{o}{=} \PY{n}{slope} \PY{o}{*} \PY{n}{x\PYZus{}end}   \PY{o}{+} \PY{n}{intercept}

\PY{k}{print} \PY{n}{x\PYZus{}base}\PY{p}{,}  \PY{n}{y\PYZus{}base}
\PY{k}{print} \PY{n}{x\PYZus{}other}\PY{p}{,} \PY{n}{y\PYZus{}other}
\PY{k}{print} \PY{n}{x\PYZus{}end}\PY{p}{,}   \PY{n}{y\PYZus{}end}
\end{Verbatim}
\textbf{Output:}
\begin{lstlisting}
0 1
1 6
2 11
\end{lstlisting}

\subsubsection{whitespace - zigzag}
\begin{Verbatim}[commandchars=\\\{\},numbers=left,firstnumber=1,stepnumber=1]
\PY{n}{intercept} \PY{o}{=} \PY{l+m+mi}{1}
\PY{n}{slope} \PY{o}{=} \PY{l+m+mi}{5}

\PY{n}{x\PYZus{}base} \PY{o}{=} \PY{l+m+mi}{0}
\PY{n}{x\PYZus{}other} \PY{o}{=} \PY{n}{x\PYZus{}base} \PY{o}{+} \PY{l+m+mi}{1}
\PY{n}{x\PYZus{}end} \PY{o}{=} \PY{n}{x\PYZus{}base} \PY{o}{+} \PY{n}{x\PYZus{}other} \PY{o}{+} \PY{l+m+mi}{1}

\PY{n}{y\PYZus{}base} \PY{o}{=} \PY{n}{slope} \PY{o}{*} \PY{n}{x\PYZus{}base} \PY{o}{+} \PY{n}{intercept}
\PY{n}{y\PYZus{}other} \PY{o}{=} \PY{n}{slope} \PY{o}{*} \PY{n}{x\PYZus{}other} \PY{o}{+} \PY{n}{intercept}
\PY{n}{y\PYZus{}end} \PY{o}{=} \PY{n}{slope} \PY{o}{*} \PY{n}{x\PYZus{}end} \PY{o}{+} \PY{n}{intercept}

\PY{k}{print} \PY{n}{x\PYZus{}base}\PY{p}{,} \PY{n}{y\PYZus{}base}
\PY{k}{print} \PY{n}{x\PYZus{}other}\PY{p}{,} \PY{n}{y\PYZus{}other}
\PY{k}{print} \PY{n}{x\PYZus{}end}\PY{p}{,} \PY{n}{y\PYZus{}end}
\end{Verbatim}
\textbf{Output:}
\begin{lstlisting}
0 1
1 6
2 11
\end{lstlisting}

\newpage
\subsection{Tables}

\begin{table*}[!ht]
\scriptsize
\renewcommand{\arraystretch}{1.2}
\begin{center}
\caption{Results by program version. (*) = log.\ regression reference, LOC = lines of code, CC = cyclomatic complexity, RT = response time. Main effects listed for version and experience. CE = prob.\ of common error, GR = grade, * = significance.}
\label{fig:results}
\vspace{0.12in}
\begin{tabular}{l|l|l|l|l|l|l|l|l}
Type & Version & LOC & CC & Avg. Grade & Mean RT (s) & Effects (ver) & (exp) & (ver $\times$ exp) \\
\hline
between & functions (*) & 24 & 7 & 4.7 & 142.8 &  & \multirow{2}{*}{CE~$\uparrow$~*} & \\
 & inline & 19 & 7 & 5.8 & 151.5 &  &  & \\
\hline
counting & nospace (*) & 3 & 2 & 8.8 & 66.6 &  &  & \\
 & twospaces & 5 & 2 & 5.9 & 55.6 & CE~$\uparrow$~*** &  & \\
\hline
funcall & nospace (*) & 4 & 2 & 9.1 & 38.6 &  &  & \\
 & space & 4 & 2 & 8.8 & 35.9 &  &  & \\
 & vars & 7 & 2 & 9.8 & 36.9 &  &  & \\
\hline
initvar & bothbad (*) & 9 & 3 & 8.7 & 63.2 &  & & \\
 & good & 9 & 3 & 7.1 & 66.0 & GR~$\downarrow$~* &  & \\
 & onebad & 9 & 3 & 7.7 & 61.6 &  &  & \\
\hline
order & inorder (*) & 14 & 4 & 8.7 & 61.4 &  &  & \\
 & shuffled & 14 & 4 & 9.1 & 68.1 & RT~$\uparrow$~* &  & RT~$\downarrow$~* \\
\hline
overload & multmixed (*) & 11 & 1 & 8.9 & 37.3 &  &  & \\
 & plusmixed & 11 & 1 & 8.7 & 41.6 & &  & RT~$\uparrow$~** \\
 & strings & 11 & 1 & 8.5 & 39.6 &  &  & \\
\hline
partition & balanced (*) & 5 & 4 & 6.9 & 45.9 &  &  & \\
 & unbalanced & 5 & 4 & 8.0 & 41.6 & CE~$\downarrow$~* &  & \\
 & unbalanced\_pivot & 6 & 4 & 8.1 & 39.6 &  &  & \\
\hline
rectangle & basic (*) & 18 & 2 & 9.7 & 76.5 &  &  & \\
 & class & 21 & 5 & 9.4 & 72.5 &  &  & \\
 & tuples & 14 & 2 & 9.5 & 80.1 & RT~$\uparrow$~** &  & RT~$\downarrow$~** \\
\hline
scope & diffname (*) & 12 & 3 & 7.2 & 58.0 &  & \multirow{2}{*}{CE~$\downarrow$~*} & \\
 & samename & 12 & 3 & 6.7 & 57.9 &  &  & \\
\hline
whitespace & linedup (*) & 14 & 1 & 8.7 & 111.7 &  &  & \\
 & zigzag & 14 & 1 & 8.5 & 108.4 & CE~$\uparrow$~* &  & \\
\hline
\end{tabular}
\end{center}
\end{table*}

\end{document}